\documentclass[twocolumn,twocolappendix]{aastex63}
\usepackage{bm}
\usepackage{amsmath,verbatim}
\usepackage{gensymb}
\usepackage{appendix}

\def\ompt{\omega_{\rm p}t}
\newcommand{\sect}[1]{Sect.~\ref{sect:#1}}

\newcommand{\figg}[1]{Fig.~\ref{fig:#1}}

\newcommand{\comp}{c/\omega_{\rm p}}

\received{xx}
\revised{xx}
\accepted{xx}
\submitjournal{ApJL}

\graphicspath{{./}{figures/}}

\begin{document}

\title{Reconnection-driven particle acceleration in relativistic shear flows}

\author[0000-0002-5951-0756]{Lorenzo Sironi}
\email{lsironi@astro.columbia.edu}
\affiliation{Department of Astronomy and Columbia Astrophysics Laboratory, Columbia University, New York, NY 10027, USA}
\author[0000-0002-5951-0756]{Michael E. Rowan}
\email{mrowan@lbl.gov}
\affiliation{Lawrence Berkeley National Laboratory, Berkeley, CA 94720, USA}
\author[0000-0002-5951-0756]{Ramesh Narayan}
\email{rnarayan@cfa.harvard.edu}
\affiliation{Harvard-Smithsonian Center for Astrophysics, Cambridge, MA 02138, USA}



\begin{abstract}
Particle energization in shear flows is  invoked to explain non-thermal emission from the boundaries of relativistic astrophysical jets. Yet, the physics of particle injection, i.e., the mechanism that allows thermal particles to
participate in shear-driven acceleration, remains unknown. With particle-in-cell simulations, we study the development of Kelvin-Helmholtz (KH) instabilities seeded by the velocity shear between a relativistic magnetically-dominated electron-positron jet and a weakly magnetized electron-ion ambient plasma. We show that, in their nonlinear stages, KH vortices generate kinetic-scale reconnection layers, which efficiently energize the jet particles, thus providing a first-principles mechanism for particle injection into shear-driven acceleration. Our work lends support to spine-sheath models of jet emission --- with a fast core/spine surrounded by a slower sheath --- and can explain the origin of radio-emitting electrons at the boundaries of relativistic jets.
\end{abstract}

\section{Introduction}\label{sect:intro}
Shear flows are ubiquitous in space and astrophysical plasmas. The free energy of the velocity shear is often invoked to accelerate charged particles to non-thermal energies \citep[e.g.,][]{rieger_19} --- via a mechanism akin to the Fermi process in converging flows \citep{fermi_49}. Shear-driven  acceleration relies on particles scattering in between regions that move toward each other because of the velocity shear. This results in a secular energy gain, as long as the particle mean free path is sufficiently long to sample a significant velocity gradient. 
In fact, the major unknown of shear-driven acceleration models is the so-called ``injection stage,'' i.e., the mechanism(s) to promote thermal particles --- that cannot participate in shear acceleration, due to their short mean free path --- to non-thermal energies.

Shear layers may be prone to the Kelvin-Helmholtz instability (KHI), driven by the transfer of momentum across the shear interface. The KHI has been thoroughly studied with linear stability analysis \citep[e.g.,][]{Blumen_75, Ferrari_78,Ferrari_80, Sharma_98, Komissarov_99, Bodo_04, Osmanov_08, Prajapati_10, Sobacchi_18, Berlok_19} and fluid-type simulations, including relativistic effects and magnetic fields \citep[e.g.,][]{Keppens_99,Ryu_00,Zhang_09, Hamlin_13, Millas_17}.
In shear layers with flow-aligned magnetic fields, the KHI opens a new possibility for dissipation: in addition to feeding off the free energy from the velocity shear, KH vortices can wrap up the field lines onto themselves, leading to dissipation via reconnection (\citealt{Faganello_08_B, Nakamura_08, Faganello_10, Faganello_12, Henri_13, Nakamura_13, Nakamura_14, Fadanelli_18}; see also \citealt{Tolman_18}, for reconnection in KHI-stable shear flows). 

In this Letter, we employ fully-kinetic particle-in-cell (PIC) simulations to demonstrate that particles are efficiently accelerated at reconnecting current sheets that are self-consistently generated by the nonlinear stages of the KHI \citep{Faganello_08_B, Faganello_12}. Our study is motivated by the limb-brightened appearance of relativistic jets, e.g., in Cygnus A \citep{boccardi_16} and M87 \citep{walker_18}.
Instabilities at relativistic jet boundaries are seen in general-relativistic magnetohydrodynamic (MHD) simulations \citep{chatterjee_19}, which however cannot probe the physics of particle acceleration.
Our work provides a first-principles mechanism for particle injection into shear-driven acceleration in relativistic magnetically-dominated jets. This lends support to spine-sheath models of jet emission \citep{sikora_16b} and can explain the origin of radio-emitting electrons at the boundaries of relativistic jets.


\section{Numerical Method and Setup}\label{sect:setup}
We perform \emph{ab initio} PIC simulations with TRISTAN-MP \citep{buneman_93,spitkovsky_05}. 
We conduct two-dimensional (2D) simulations in the $xy$ plane, retaining all three components of particle velocities and electromagnetic fields.  In the initial state, the fluid bulk motion is along $y$, and the gradient of the velocity is along $x$. The  domain has length $L_y$ along $y$ ($0\leq y/L_y\leq1$), and width $L_x = 3L_y$ along $x$ ($-1.5\leq x/L_y\leq1.5$), with periodic boundary conditions in both directions. We initialize a relativistic magnetically-dominated jet at $|x|/L_x\lesssim 0.25$, and a weakly magnetized stationary ambient plasma at  $|x|/L_x\gtrsim 0.25$ (which we call ``wind,'' since it should represent the wind of the accretion flow). The simulation is performed in the wind frame.

The jet is composed of electron-positron pairs with comoving density $n_0$ (including both species) and a small thermal spread, moving with four-velocity $\Gamma_0\beta_0=1.3$ along $+\hat{y}$ (we also report results for $\Gamma_0\beta_0=3$ and 10). The jet carries an energetically-dominant magnetic field. The in-plane field strength $B_{\textrm{j},y}$ is parameterized by the magnetization $\sigma_{\textrm{j},y}=B_{\textrm{j},y}^2/(4\pi n_0 m_{\rm e} c^2)$, where $m_{\rm e}$ is the electron mass and $c$ the speed of light. We also initialize an out-of-plane field $B_{\textrm{j},z}\equiv B_{z0}=B_{\textrm{j},y}\tan\theta$ and its associated motional electric field $E_{\textrm{j},x}=-\beta_0 B_{\textrm{j},z}$. Our reference runs employ $\sigma_{\textrm{j},y}=6.7$ and $\theta=75^\circ$ (corresponding to $\theta'=65^\circ$ in the jet frame for $\Gamma_0\beta_0=1.3$), since astrophysical jets are magnetically-dominated and have comparable poloidal (here, along $y$) and toroidal (along $z$) fields in the jet frame (see, e.g., \citealt{Alves_14,Alves_15,Liang_13a,Liang_13b,Nishikawa_14,Nishikawa_16,Pausch_17} for PIC studies of shear instabilities in unmagnetized plasmas).

The wind is composed of an electron-ion plasma with density $n_{\rm w}=128\,n_0$ (including both species). We typically employ a mass ratio $m_{\rm i}/m_{\rm e}=25$, but we have tested that the late-time particle spectrum is the same for $m_{\rm i}/m_{\rm e}=5$, 25, and 100, and also for the artificial case of an electron-positron wind ($m_{\rm i}/m_{\rm e}=1$). As we show below, particle energization primarily involves the jet particles, and so our results are insensitive to the mass ratio in the wind. The wind is initialized with an out-of-plane field $B_{\textrm{w},z}=0.1\,B_{\textrm{j},z}$. The wind plasma beta $\beta_{\rm p}\approx [B_{\textrm{j},y}^2+(\Gamma_0^{-1}B_{\textrm{j},z})^2]/B_{\textrm{w},z}^2\approx 40\gg1$, so the wind is particle-dominated.

Our unit of length is the skin depth of jet particles, $\comp=\sqrt{m_{\rm e} c^2/(4\pi e^2 n_0)}$, which we resolve with 11.3 cells ($e$ is the positron charge). The electron skin depth and Debye length in the wind 
are marginally resolved. Our reference runs have $L_y\equiv L_0=3840\,{\rm cells}\approx340\,\comp$, but we also present larger runs with $L_y=3L_0\approx 1020 \,\comp$ to demonstrate that we achieve asymptotically-converged results. 

The wind and jet properties are smoothly connected with spatial profiles varying as $\tanh[2\pi (x-x_{\rm SL})/\Delta]$ in the vicinity of the shear layers at $|x|=x_{\rm SL}\approx 0.25\,L_x$. Our emphasis is on wide shear layers (with thickness $\Delta \gg \comp$), in application to realistic jet/wind boundaries. For $\Delta \gg \comp$, the  KHI growth should be independent of kinetic physics, and in fact our measured growth rates are in good agreement with MHD expectations \citep[e.g.,][]{Bodo_04}. To ensure that we start from MHD-scale initial conditions, we choose $\Delta$ to be larger than the largest kinetic scale, i.e., the  Larmor radius of wind ions, $r_{\rm L,i}\approx(\Gamma_0^{-1}B_{\textrm{j},z}/B_{\textrm{w},z})\sqrt{(m_{\rm i}/m_{\rm e})(n_0/n_{\rm w})}\,\comp$. We typically employ $\Delta=64\,\comp$, but we report identical results obtained with $\Delta=192\,\comp$. The spatial profiles of temperature, charge density and electric current density in the shear layer follow from pressure equilibrium and Maxwell's equations.

We employ 4 particles per cell in the jet. For computational convenience, in the wind we typically use particles with a larger numerical weight (fixing the overall wind mass density, this gives a lower macro-particle count), but we have carefully checked that our results are insensitive to this choice. 

\begin{figure}
    \centering
    \includegraphics[width=0.5\textwidth]{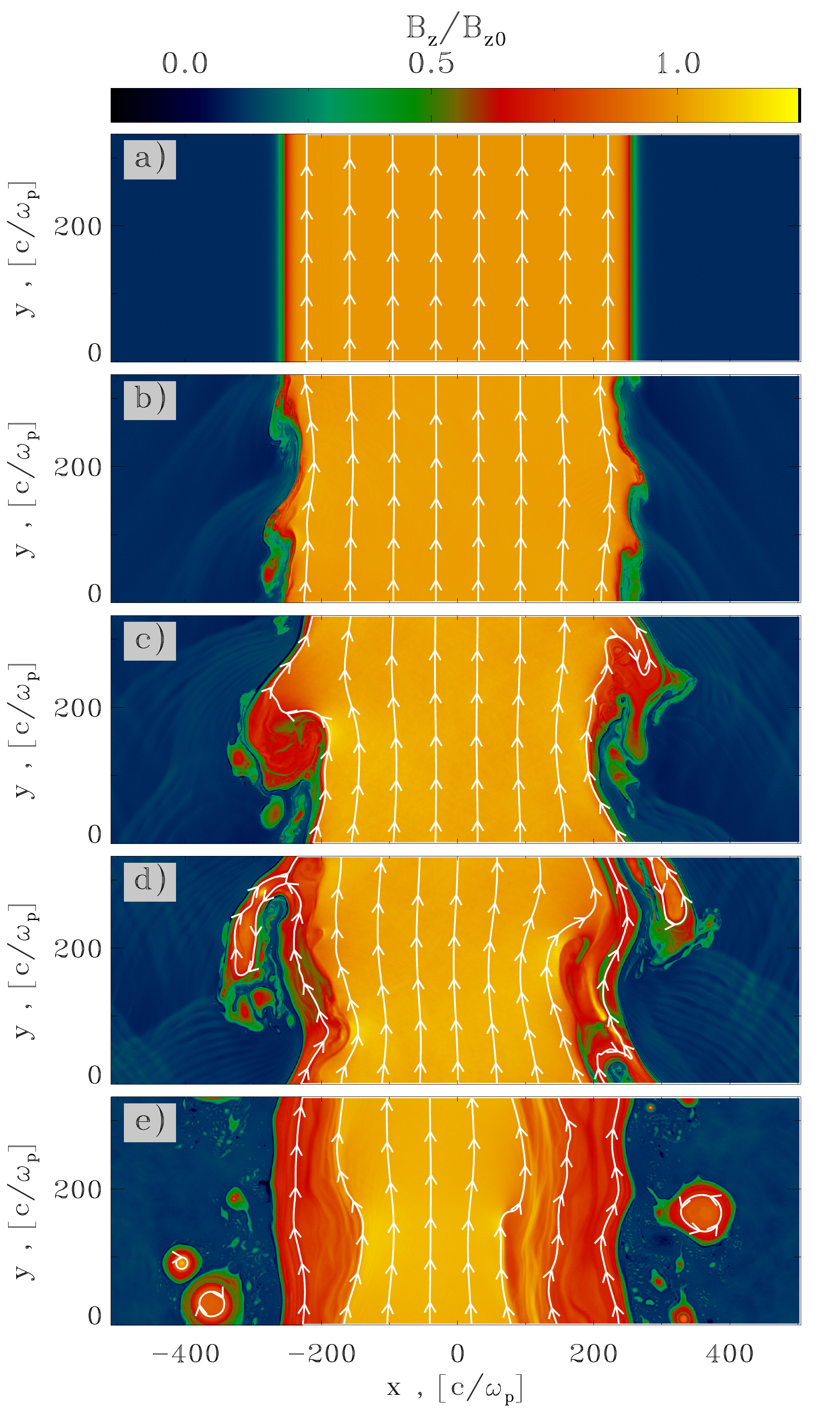}
    \caption{2D evolution of the out-of-plane field $B_z$ (color scale) --- in units of the initial jet field $B_{z0}\equiv B_{\textrm{j},z}$ --- at (a) $\ompt=80$, (b) $\ompt=3262$, (c) $\ompt=4216$, (d) $\ompt=5171$, and (e) $\ompt=12569$, with in-plane field lines overlaid. The magnetized jet is initially at $|x|\lesssim 250\,\comp$, surrounded by the wind.}
    \label{fig:fluidtime}
\end{figure}

\begin{figure}
    \centering
    \includegraphics[width=0.5\textwidth]{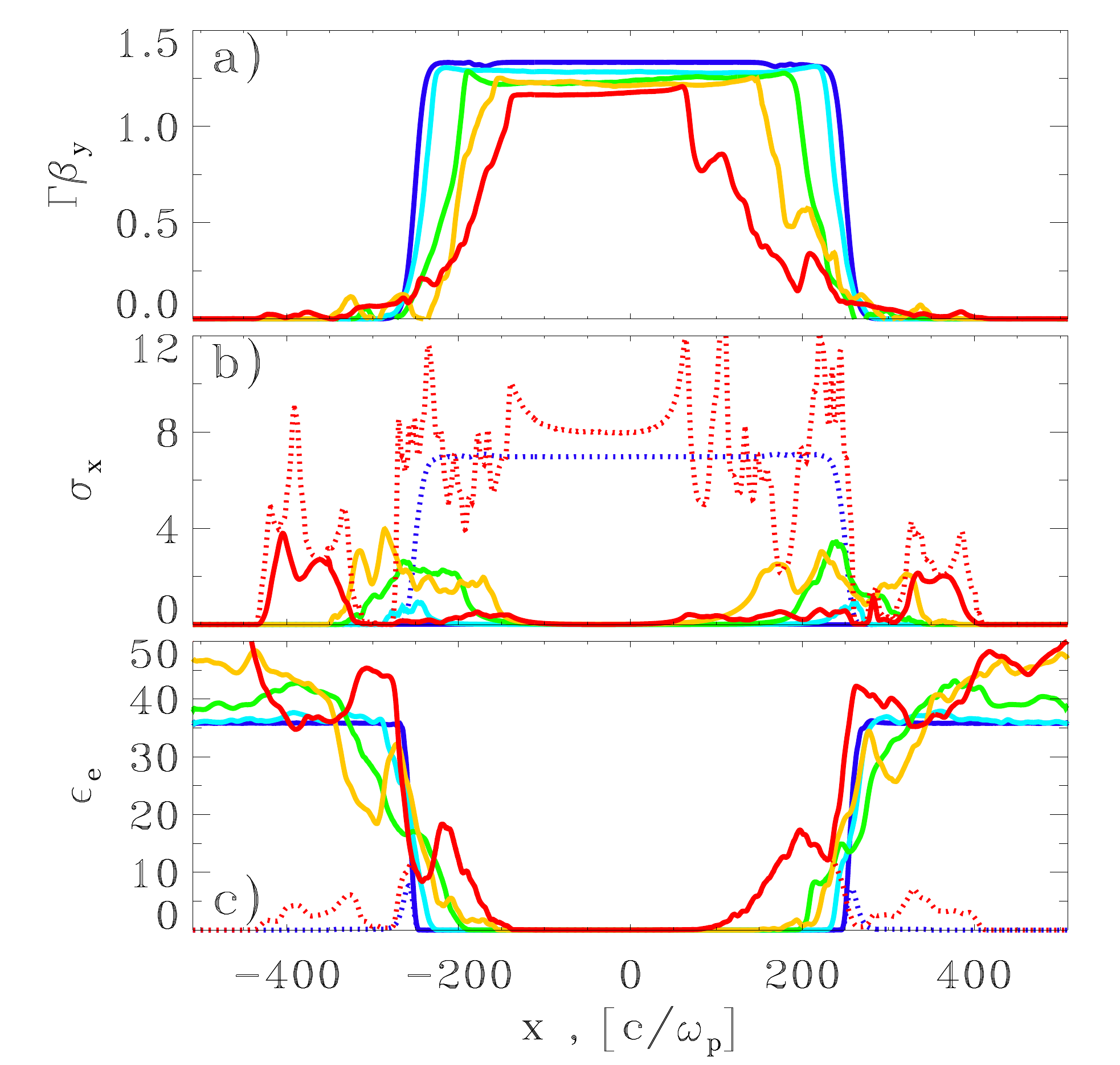}
    \caption{Temporal evolution of $y$-averaged profiles, with colors from blue to red referring to the same times as panels in \figg{fluidtime}. (a) Bulk four-velocity along $y$, in units of the speed of light, where in each cell the fluid speed is computed by averaging over velocities of individual particles. (b) Local magnetization contributed by $B_x$ (solid lines), i.e., $\sigma_{x}=B_x^2/(4\pi n_0 m_{\rm e} c^2)$. Dotted lines represent the magnetization from in-plane fields, i.e., $\sigma_{x}+\sigma_{y}$, at the initial (blue) and final (red) times. (c) Electron internal energy density normalized to the initial rest-mass energy density of jet electrons. Dotted lines refer to jet electrons only, at the initial (blue) and final (red) times.} 
    \label{fig:fluidprof}
\end{figure}

\section{Results}\label{sect:results}
The temporal evolution of the KHI is presented in \figg{fluidtime}, where color indicates the out-of-plane field $B_z$, with in-plane field lines overlaid. The instability develops in two stages: a mode with wavelength $\lambda\approx L_y/2$ appears in panel (b), whereas a longer-wavelength mode with $\lambda\approx L_y$ grows at later times. The corresponding growth rates are in good agreement with MHD linear dispersion analysis. The vortices created by the nonlinear stages of the KHI bend the in-plane field lines, creating anti-parallel configurations prone to reconnection (panel (d)). The final stage (panel (e)) is characterized by: (\textit{i}) the persistence of a nearly-unperturbed jet core (yellow) surrounded by a sheath of weaker $B_z$ (red), whose width is ${\approx} 0.3\,L_y$; (\textit{ii}) the presence of magnetized ``clouds'' of jet material --- on a variety of scales, from plasma scales up to ${\approx} 0.3\,L_y$ --- in pressure equilibrium with the surrounding wind plasma. 

The evolution of the KHI is further analyzed in \figg{fluidprof}.
The jet starts with bulk four-velocity $\Gamma\beta_y = \Gamma_0\beta_0 = 1.3$ (blue in (a)), and it is magnetically-dominated, with $\sigma_{y}\approx\sigma_{{\rm j},y}=6.7$ (dotted blue in (b)) and $\sigma_{z}= \sigma_{y}\tan^2\theta\,\approx 93.3$ (here, $\sigma_{i}\equiv B_i^2/4\pi n_0 m_{\rm e} c^2$ is the magnetization contributed by the field component $B_i$). As a result of the KHI, the in-plane field lines are twisted and folded, and 
a significant $B_x$ develops at the jet boundaries, with peak magnetization $\sigma_{x}\approx4$ (green and yellow in (b)). Since part of the resulting magnetic energy will be dissipated by reconnection, the peak value of $\sigma_{x}$ can be taken as a proxy for the characteristic magnetization of reconnecting current sheets. Since $\sigma_{x}\gtrsim 1$, KHI-driven reconnection occurs in the relativistic regime.

The end stage (red lines) shows a velocity profile characterized by a fast jet core (at $|x|\lesssim 100\,\comp$), moving nearly at the initial four-velocity $\Gamma_0\beta_0 = 1.3$, surrounded by wings (or, a sheath) of slower moving material (at $100\lesssim |x|\lesssim 250\,\comp$), with $\Gamma \beta_y\approx 0.5$. A trans-relativistic sheath also characterizes the final stages of simulations starting with faster jets ($\Gamma_0\beta_0 = 3$ and $\Gamma_0\beta_0 = 10$). In the sheath near $|x|\approx 200\,\comp$, the in-plane magnetic energy density (dotted red in (b)) is nearly in equipartition with the electron energy density (solid red in (c)), or equivalently, the plasma beta $\beta_{\rm p}\approx 1$. This is a generic outcome of relativistic reconnection \citep[e.g.,][]{sironi_16}.

\begin{figure}
    \centering
    \includegraphics[width=0.5\textwidth]{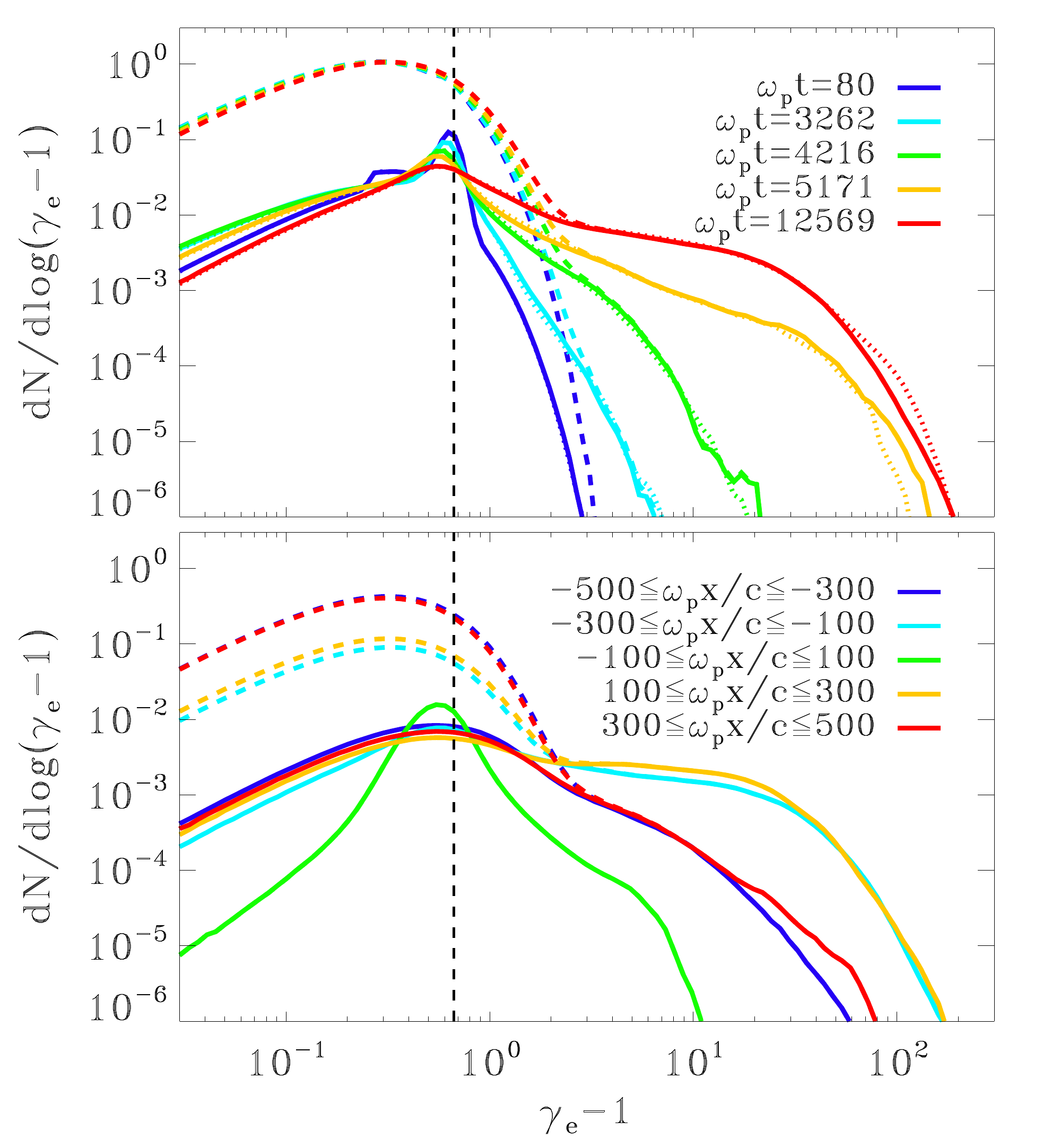}
    \caption{Top: Temporal evolution of the electron spectrum (at the times indicated in the legend, corresponding to the panels in \figg{fluidtime}), for all the electrons (dashed), jet electrons only (solid), and jet positrons (dotted). Bottom: at the final time $\ompt=12569$, spatial dependence of the electron  spectrum (see legend), for all the electrons (dashed) and jet electrons only (solid). In both panels, the vertical black dashed line is at the bulk energy $\Gamma_0-1$.}
    \label{fig:spectime}
\end{figure}

The nonlinear development of the KHI leads to efficient particle acceleration (the temporal evolution of the electron spectrum is in the top panel of \figg{spectime}). Wind electrons populate a non-relativistic Maxwellian (dashed lines), while the spectrum of jet electrons initially peaks at their bulk energy $\Gamma_0-1\approx 0.7$ (solid blue). Concurrently with the formation of KHI-induced current sheets, a distinct high-energy component emerges, primarily populated by jet particles (green and yellow correspond to the times of \figg{fluidtime}(c) and (d)). The spectral cutoff shifts up in energy at every stage of nonlinear KHI development (i.e., first with the $\lambda\approx L_y/2$ mode going nonlinear, and then with the $\lambda\approx L_y$ mode).
In the final stage (red lines), the spectrum extends up to a cutoff energy $\gamma_{\rm e}-1\approx 30$, as expected from reconnection-driven particle acceleration if the in-plane magnetization $\approx 10$ \citep{werner_16,petropoulou_18}, as inferred from \figg{fluidprof}(b). The electron spectrum at even later times (not shown) is nearly identical to the red curve, i.e., the system has reached a a steady state. At all times, the spectrum of jet positrons (dotted) is nearly identical to the one of jet electrons (solid).

Most of the electron and positron acceleration is localized at the jet boundaries, with nearly identical outcomes from the left and right side (bottom panel in \figg{spectime}, cyan and yellow lines). The core of the jet (green line) retains a narrow spectrum centered at the bulk energy $\Gamma_0-1\approx 0.7$. The high-energy particles at $|x|\gtrsim 300\,\comp$ (blue and red lines) were initially in the jet, and now they reside in the magnetized clouds embedded in the wind (\figg{fluidtime}(e)).

\begin{figure}
    \centering
    \includegraphics[width=0.5\textwidth]{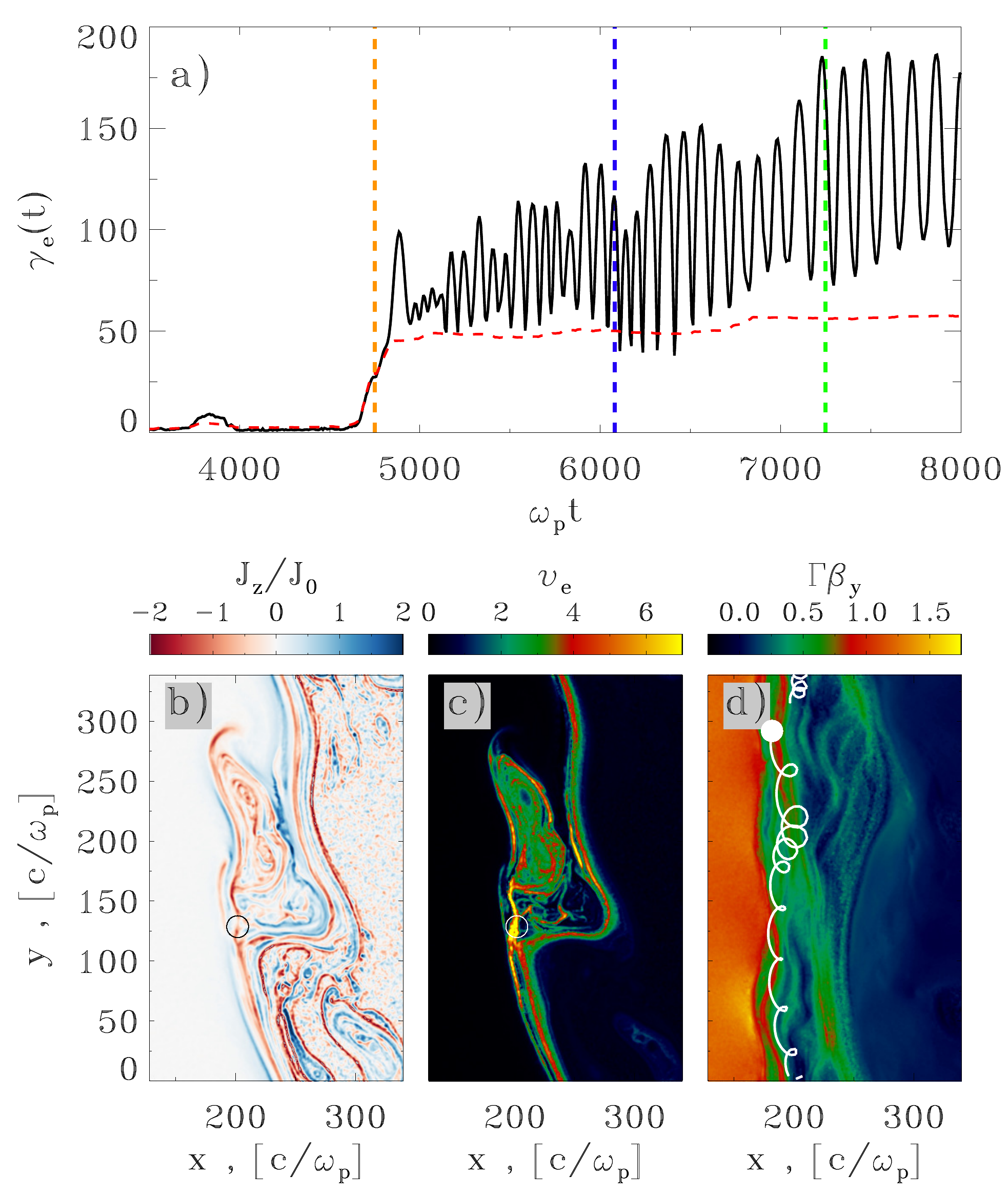}
    \caption{Trajectory of a representative high-energy electron. (a) Time evolution of the electron Lorentz factor (black solid) and of the parallel work $W_\parallel=-e\int_0^t E_\parallel v_\parallel dt'/m_{\rm e}c^2$ (dashed red), where $E_\parallel=\it{\boldsymbol{E}}\cdot \hat{\it{\boldsymbol{b}}}$ and $v_\parallel=\it{\boldsymbol{v}}\cdot \hat{\it{\boldsymbol{b}}}$ (here, $\it{\boldsymbol{E}}$ is the electric field, $\it{\boldsymbol{v}}$  the electron velocity and $\hat{\it{\boldsymbol{b}}}={\it{\boldsymbol{B}}}/B$  the magnetic field unit vector). (b) and (c): 2D structure of the out-of-plane current $J_z$ (in units of $J_0=en_0 c$) and of the mean internal energy per electron $\upsilon_{\rm e}$ (in units of $m_{\rm e} c^2$), at the time $\ompt=4750$ of particle injection (vertical dashed orange in panel (a)). The electron position at this time is indicated by the circle. (d) 2D structure of the bulk four-velocity along $y$, in units of the speed of light, at $\ompt\approx 7250$ (vertical dashed green line in panel (a)). The electron position at this time is indicated by the filled white circle, and we also plot its trajectory from $\ompt=6080$ (blue dashed in panel (a)) to $\ompt=7250$  (green dashed in (a)).} 
    \label{fig:testprt}
\end{figure}

The trajectory of a representative high-energy electron is displayed in \figg{testprt}. The top panel shows that the first stage of particle acceleration ($\ompt\approx 4750$, marked by the vertical orange line) is powered by $E_\parallel=\it{\boldsymbol{E}}\cdot \hat{\it{\boldsymbol{b}}}$ (dashed red in (a)). 
This is indeed expected for reconnection-powered acceleration with a strong non-alternating component \citep{ball_sironi_19,comisso_19b}. During this injection stage, the electron is located within a reconnecting current sheet (panel (b)), where efficient particle acceleration/heating occurs (panel (c)). At later times, while $E_\parallel$ no longer results in acceleration, the electron energy still steadily grows --- a similar two-stage acceleration process has been reported for  magnetically-dominated plasma turbulence \citep{comisso_18,comisso_19b}. In this time range (between the vertical dashed blue and green lines in (a)), the electron gains energy while moving back and forth across the shear layer (panel (d)), as expected in shear-driven acceleration. At this stage, the electron orbit covers a sizeable fraction of the shear layer width, and so it can experience a significant velocity gradient.

\begin{figure}
    \centering
    \includegraphics[width=0.5\textwidth]{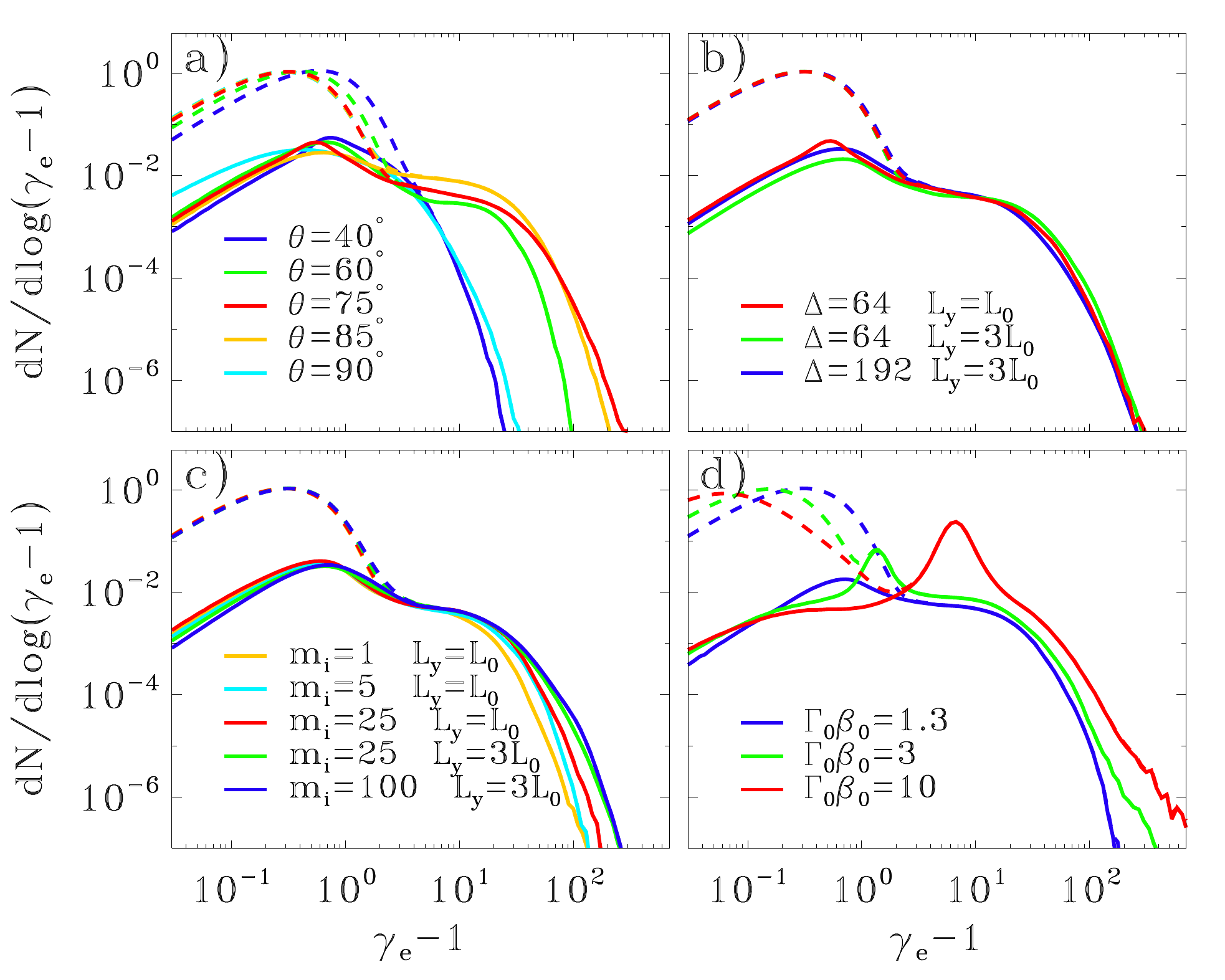}
    \caption{Dependence of the electron spectrum on physical parameters, for all the electrons (dashed) and jet electrons only (solid). Unless otherwise noted, we employ numerical and physical parameters as described in \sect{setup} for our reference run. (a) We vary the simulation-frame angle $\theta=\arctan(B_{\textrm{j},z}/B_{\textrm{j},y})$ as indicated in the legend, keeping the initial $B_{\textrm{j},y}^2+B_{\textrm{j},z}^2$ fixed. Spectra refer to $\ompt=12410$. (b) We vary the layer width $\Delta$ (in units of $\comp$) and the box size $L_y$ (see legend). Spectra refer to $t\approx 31\,L_y/c$, corresponding to $\ompt=10500$ for $L_y=L_0$ and to $\ompt=31500$ for $L_y=3\,L_0$. (c) We vary the ion mass $m_{\rm i}$ (in units of $m_{\rm e}$, see legend). Yellow, cyan, and red spectra refer to our reference box at $\ompt=11137$, while green and blue spectra refer to a box with wider $\Delta=192\,\comp$ and larger size $L_y=3L_0$, and they are measured at $\ompt=33411$ (so, all spectra are measured at $t\approx 33\,L_y/c$). (d) We vary the jet bulk four-velocity (see legend). Spectra refer to simulations with $m_{\rm i}/m_{\rm e}=1$ and $\Delta=16\,\comp$ at $\ompt=13523$.}
    \label{fig:speccomp}
\end{figure}

To assess the generality of reconnection-powered injection in KHI-unstable shear layers, in \figg{speccomp} we present the dependence of the spectrum of all electrons (dashed) and jet electrons (solid) on several physical parameters. Spectra are shown at sufficiently late times that the system is nearly in steady state. 
When varying the lab-frame angle $\theta=\arctan(B_{\textrm{j},z}/B_{\textrm{j},y})$ at fixed $B_{\textrm{j},y}^2+B_{\textrm{j},z}^2$ (panel (a)), we find that reconnection-driven particle acceleration is most efficient at intermediate angles, $60^\circ\lesssim\theta\lesssim85^\circ$. At smaller angles ($\theta=40^\circ$), the shear layer is KHI-stable. 
In the absence of in-plane fields ($\theta=90^\circ$), reconnection cannot operate, and we report only  marginal evidence for accelerated particles, with cutoff energy much smaller than in our reference run (see also \citealt{Cerutti_20}). 
The high-energy spectral cutoff does not significantly vary for angles $60^\circ\lesssim\theta\lesssim85^\circ$. This is a consequence of the fact that the typical magnetization of KHI-generated current sheets (as tracked by the peak $\sigma_{x}$) is nearly the same, for $\theta$ in this range. In turn, this is due to a combination of two opposite effects: at larger $\theta$, the initial $\sigma_{\textrm{j},y}$ is smaller, yet the KHI is more effective in folding the field lines (precisely because of the weaker tension of in-plane fields), which results, overall, in comparable magnetizations of the self-generated current sheets. Given that black-hole jets start with poloidal fields (here, along $y$), while they are dominated by toroidal fields (here, along $z$) at large distances, our results in \figg{speccomp}(a) may help put constraints on the distance where KHI-driven reconnection and ensuing particle acceleration is most effective.

We have also tested the dependence of our steady-state spectra on the shear layer width $\Delta$ and the box size $L_y$, demonstrating that our results hold in the MHD limit $L_y\gg\comp$ and $\Delta\gg\comp$ (panel (b)). Electron spectra also show only a weak dependence on the ion-to-electron mass ratio in the wind (panel (c)); this is not surprising, given that particle acceleration mostly involves  electrons and positrons in the jet. 

Finally, panel (d) illustrates the dependence on the initial jet velocity: with increasing $\Gamma_0\beta_0$, a separate population emerges at high energies ($\gamma_{\rm e}\gtrsim 200$), beyond the bump of reconnection-accelerated particles at $\gamma_{\rm e}\lesssim 100$. By tracking the  trajectories of the highest-energy electrons reaching $\gamma_{\rm e}\gtrsim 200$, we find that they are accelerated first by reconnection, and then by scattering back and forth across the shear layer (as in \figg{testprt}), i.e., they participate in shear-driven acceleration.


\section{Conclusions}\label{sect:concl}
By means of large-scale 2D PIC simulations, we have studied the physics of particle acceleration in KHI-unstable shear layers, for the conditions expected at the boundary of relativistic magnetically-dominated jets. We start from shear layers much wider than kinetic plasma scales. We find that the nonlinear evolution of KH vortices leads to reconnection of the jet magnetic field, which results in efficient acceleration of jet electrons and positrons. The highest energy particles resulting from reconnection are further energized by shear-driven acceleration, i.e., reconnection can mediate particle injection into shear acceleration. Our work lends support to spine-sheath models of jet emission
and can explain the origin of radio-emitting electrons at the boundaries of relativistic jets (see \citealt{ripperda_20} for an alternative explanation).

We defer an investigation of 3D effects to future work, though we note that simulations of both relativistic reconnection \citep[e.g.][]{ss_14,guo_14,werner_17,sironi_belo_20} and magnetically-dominated plasma turbulence \citep[e.g.,][]{comisso_18,comisso_19b} yield similar results between 2D and 3D, so we expect our conclusions to be applicable to the 3D case. We also leave to future work a more detailed characterization of the properties of shear-accelerated particles (e.g., acceleration efficiency, power-law slope, scattering and acceleration rates).

\acknowledgments
This work is supported in part by NASA via the TCAN
award grant NNX14AB47G and by the black hole initiative at
Harvard University, which is supported by a grant from
the Templeton Foundation.
LS acknowledges support from the Sloan Fellowship, the Cottrell Scholar Award, NASA ATP NNX17AG21G and NSF PHY-1903412. The simulations have been performed at Columbia (Habanero and Terremoto), and with NASA (Pleiades) resources.

\appendix
\section{KHI-driven reconnection plasmoids}
\begin{figure}
    \centering
    \includegraphics[width=0.5\textwidth]{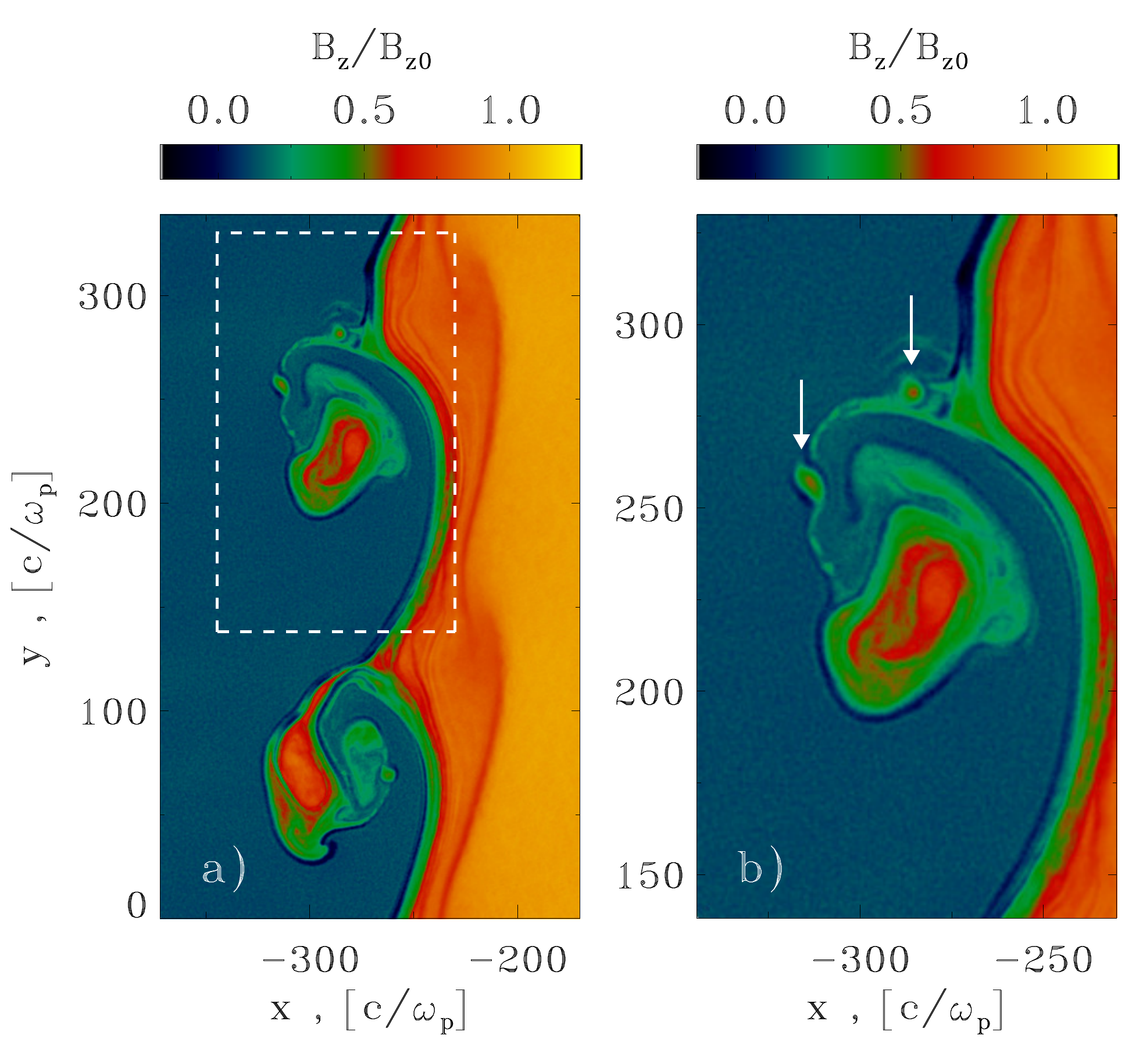}
    \caption{2D structure of the 
    out-of-plane field $B_z$, in units of $B_{z0}\equiv B_{{\rm j,z}}$, from a simulation with $\theta=60^\circ$, whose late-time spectrum is shown by the green line in \figg{speccomp}(a). Both panels refer to $\ompt=3740$, when the $\lambda=L_y/2$ mode goes nonlinear. The right panel is a zoom-in view of the left panel (from the area delimited by the white dashed lines), and it shows the presence of two reconnection plasmoids (as marked by the white arrows).} 
    \label{fig:plasmoids}
\end{figure}
In the main body of the paper we have demonstrated that the nonlinear development of the KHI naturally produces reconnection current sheets, which are conducive to efficient particle acceleration. Long reconnection layers are known to be prone to the tearing mode instability \citep[e.g.,][]{uzdensky_10,huang_12,loureiro_12}, which breaks the current sheet into a chain of plasmoids/flux ropes. In \figg{plasmoids}, we show that this indeed occurs for KHI-generated reconnection layers: the KH vortex in the upper half of the left panel displays two reconnection plasmoids (marked by the arrows in the zoom-in view on the right panel). We argue that reconnection plasmoids, together with nonlinear structures generated by the KHI, can provide the scattering required for efficient shear-driven acceleration.

\bibliography{blob_ApJL,kh_refs,blob,extra,NASA}{}

\begin{thebibliography}{}
\expandafter\ifx\csname natexlab\endcsname\relax\def\natexlab#1{#1}\fi
\providecommand{\url}[1]{\href{#1}{#1}}
\providecommand{\dodoi}[1]{doi:~\href{http://doi.org/#1}{\nolinkurl{#1}}}
\providecommand{\doeprint}[1]{\href{http://ascl.net/#1}{\nolinkurl{http://ascl.net/#1}}}
\providecommand{\doarXiv}[1]{\href{https://arxiv.org/abs/#1}{\nolinkurl{https://arxiv.org/abs/#1}}}

\bibitem[{{Alves} {et~al.}(2014){Alves}, {Grismayer}, {Fonseca}, \&
  {Silva}}]{Alves_14}
{Alves}, E.~P., {Grismayer}, T., {Fonseca}, R.~A., \& {Silva}, L.~O. 2014, New
  Journal of Physics, 16, 035007, \dodoi{10.1088/1367-2630/16/3/035007}

\bibitem[{Alves {et~al.}(2015)Alves, Grismayer, Fonseca, \& Silva}]{Alves_15}
Alves, E.~P., Grismayer, T., Fonseca, R.~A., \& Silva, L.~O. 2015, Phys. Rev.
  E, 92, 021101, \dodoi{10.1103/PhysRevE.92.021101}

\bibitem[{{Ball} {et~al.}(2019){Ball}, {Sironi}, \&
  {{\"O}zel}}]{ball_sironi_19}
{Ball}, D., {Sironi}, L., \& {{\"O}zel}, F. 2019, \apj, 884, 57,
  \dodoi{10.3847/1538-4357/ab3f2e}

\bibitem[{{Berlok} \& {Pfrommer}(2019)}]{Berlok_19}
{Berlok}, T., \& {Pfrommer}, C. 2019, \mnras, 485, 908,
  \dodoi{10.1093/mnras/stz379}

\bibitem[{{Blumen} {et~al.}(1975){Blumen}, {Drazin}, \& {Billings}}]{Blumen_75}
{Blumen}, W., {Drazin}, P.~G., \& {Billings}, D.~F. 1975, Journal of Fluid
  Mechanics, 71, 305, \dodoi{10.1017/S0022112075002595}

\bibitem[{{Boccardi} {et~al.}(2016){Boccardi}, {Krichbaum}, {Bach}, {Mertens},
  {Ros}, {Alef}, \& {Zensus}}]{boccardi_16}
{Boccardi}, B., {Krichbaum}, T.~P., {Bach}, U., {et~al.} 2016, \aap, 585, A33,
  \dodoi{10.1051/0004-6361/201526985}

\bibitem[{Bodo {et~al.}(2004)Bodo, Mignone, \& Rosner}]{Bodo_04}
Bodo, G., Mignone, A., \& Rosner, R. 2004, Phys. Rev. E, 70, 036304

\bibitem[{{Buneman}(1993)}]{buneman_93}
{Buneman}, O. 1993, {in ``Computer Space Plasma Physics,'' Terra Scientific,
  Tokyo, 67}

\bibitem[{{Cerutti} \& {Giacinti}(2020)}]{Cerutti_20}
{Cerutti}, B., \& {Giacinti}, G. 2020, arXiv e-prints, arXiv:2008.07253.
\newblock \doarXiv{2008.07253}

\bibitem[{{Chatterjee} {et~al.}(2019){Chatterjee}, {Liska}, {Tchekhovskoy}, \&
  {Markoff}}]{chatterjee_19}
{Chatterjee}, K., {Liska}, M., {Tchekhovskoy}, A., \& {Markoff}, S.~B. 2019,
  \mnras, 490, 2200, \dodoi{10.1093/mnras/stz2626}

\bibitem[{{Comisso} \& {Sironi}(2018)}]{comisso_18}
{Comisso}, L., \& {Sironi}, L. 2018, \prl, 121, 255101,
  \dodoi{10.1103/PhysRevLett.121.255101}

\bibitem[{{Comisso} \& {Sironi}(2019)}]{comisso_19b}
---. 2019, \apj, 886, 122, \dodoi{10.3847/1538-4357/ab4c33}

\bibitem[{Fadanelli {et~al.}(2018)Fadanelli, Faganello, Califano, Cerri,
  Pegoraro, \& Lavraud}]{Fadanelli_18}
Fadanelli, S., Faganello, M., Califano, F., {et~al.} 2018, Journal of
  Geophysical Research: Space Physics, 123, 9340, \dodoi{10.1029/2018JA025626}

\bibitem[{Faganello {et~al.}(2008)Faganello, Califano, \&
  Pegoraro}]{Faganello_08_B}
Faganello, M., Califano, F., \& Pegoraro, F. 2008, Phys. Rev. Lett., 101,
  105001, \dodoi{10.1103/PhysRevLett.101.105001}

\bibitem[{{Faganello} {et~al.}(2012){Faganello}, {Califano}, {Pegoraro},
  {Andreussi}, \& {Benkadda}}]{Faganello_12}
{Faganello}, M., {Califano}, F., {Pegoraro}, F., {Andreussi}, T., \&
  {Benkadda}, S. 2012, Plasma Physics and Controlled Fusion, 54, 124037,
  \dodoi{10.1088/0741-3335/54/12/124037}

\bibitem[{Faganello {et~al.}(2010)Faganello, Pegoraro, Califano, \&
  Marradi}]{Faganello_10}
Faganello, M., Pegoraro, F., Califano, F., \& Marradi, L. 2010, Physics of
  Plasmas, 17, 062102, \dodoi{10.1063/1.3430640}

\bibitem[{{Fermi}(1949)}]{fermi_49}
{Fermi}, E. 1949, Physical Review, 75, 1169, \dodoi{10.1103/PhysRev.75.1169}

\bibitem[{{Ferrari} {et~al.}(1978){Ferrari}, {Trussoni}, \&
  {Zaninetti}}]{Ferrari_78}
{Ferrari}, A., {Trussoni}, E., \& {Zaninetti}, L. 1978, \aap, 64, 43

\bibitem[{{Ferrari} {et~al.}(1980){Ferrari}, {Trussoni}, \&
  {Zaninetti}}]{Ferrari_80}
---. 1980, \mnras, 193, 469, \dodoi{10.1093/mnras/193.3.469}

\bibitem[{{Guo} {et~al.}(2014){Guo}, {Li}, {Daughton}, \& {Liu}}]{guo_14}
{Guo}, F., {Li}, H., {Daughton}, W., \& {Liu}, Y.-H. 2014, Physical Review
  Letters, 113, 155005, \dodoi{10.1103/PhysRevLett.113.155005}

\bibitem[{Hamlin \& Newman(2013)}]{Hamlin_13}
Hamlin, N., \& Newman, W. 2013, Phys. Rev. E, 87, 043101

\bibitem[{{Henri} {et~al.}(2013){Henri}, {Cerri}, {Califano}, {Pegoraro},
  {Rossi}, {Faganello}, {{\v{S}}ebek}, {Tr{\'a}vn{\'\i}{\v{c}}ek}, {Hellinger},
  {Frederiksen}, {Nordlund}, {Markidis}, {Keppens}, \& {Lapenta}}]{Henri_13}
{Henri}, P., {Cerri}, S.~S., {Califano}, F., {et~al.} 2013, Physics of Plasmas,
  20, 102118, \dodoi{10.1063/1.4826214}

\bibitem[{{Huang} \& {Bhattacharjee}(2012)}]{huang_12}
{Huang}, Y.-M., \& {Bhattacharjee}, A. 2012, Physical Review Letters, 109,
  265002, \dodoi{10.1103/PhysRevLett.109.265002}

\bibitem[{Keppens {et~al.}(1999)Keppens, T{\'o}th, Westermann, \&
  Goedbloed}]{Keppens_99}
Keppens, R., T{\'o}th, G., Westermann, R., \& Goedbloed, J. 1999, Journal of
  Plasma Physics, 61, 1, \dodoi{10.1017/S0022377898007223}

\bibitem[{Komissarov(1999)}]{Komissarov_99}
Komissarov, S.~S. 1999, MNRAS, 303, 343

\bibitem[{{Liang} {et~al.}(2013{\natexlab{a}}){Liang}, {Boettcher}, \&
  {Smith}}]{Liang_13a}
{Liang}, E., {Boettcher}, M., \& {Smith}, I. 2013{\natexlab{a}}, \apjl, 766,
  L19, \dodoi{10.1088/2041-8205/766/2/L19}

\bibitem[{{Liang} {et~al.}(2013{\natexlab{b}}){Liang}, {Fu}, {Boettcher},
  {Smith}, \& {Roustazadeh}}]{Liang_13b}
{Liang}, E., {Fu}, W., {Boettcher}, M., {Smith}, I., \& {Roustazadeh}, P.
  2013{\natexlab{b}}, \apjl, 779, L27, \dodoi{10.1088/2041-8205/779/2/L27}

\bibitem[{{Loureiro} {et~al.}(2012){Loureiro}, {Samtaney}, {Schekochihin}, \&
  {Uzdensky}}]{loureiro_12}
{Loureiro}, N.~F., {Samtaney}, R., {Schekochihin}, A.~A., \& {Uzdensky}, D.~A.
  2012, Physics of Plasmas, 19, 042303, \dodoi{10.1063/1.3703318}

\bibitem[{{Millas} {et~al.}(2017){Millas}, {Keppens}, \& {Meliani}}]{Millas_17}
{Millas}, D., {Keppens}, R., \& {Meliani}, Z. 2017, \mnras, 470, 592,
  \dodoi{10.1093/mnras/stx1288}

\bibitem[{Nakamura \& Daughton(2014)}]{Nakamura_14}
Nakamura, T. K.~M., \& Daughton, W. 2014, Geophysical Research Letters, 41,
  8704, \dodoi{10.1002/2014GL061952}

\bibitem[{Nakamura {et~al.}(2013)Nakamura, Daughton, Karimabadi, \&
  Eriksson}]{Nakamura_13}
Nakamura, T. K.~M., Daughton, W., Karimabadi, H., \& Eriksson, S. 2013, Journal
  of Geophysical Research: Space Physics, 118, 5742, \dodoi{10.1002/jgra.50547}

\bibitem[{Nakamura \& Fujimoto(2008)}]{Nakamura_08}
Nakamura, T. K.~M., \& Fujimoto, M. 2008, Phys. Rev. Lett., 101, 165002,
  \dodoi{10.1103/PhysRevLett.101.165002}

\bibitem[{{Nishikawa} {et~al.}(2014){Nishikawa}, {Hardee}, {Du{\c{t}}an},
  {Niemiec}, {Medvedev}, {Mizuno}, {Meli}, {Sol}, {Zhang}, {Pohl}, \&
  {Hartmann}}]{Nishikawa_14}
{Nishikawa}, K.~I., {Hardee}, P.~E., {Du{\c{t}}an}, I., {et~al.} 2014, \apj,
  793, 60, \dodoi{10.1088/0004-637X/793/1/60}

\bibitem[{{Nishikawa} {et~al.}(2016){Nishikawa}, {Frederiksen}, {Nordlund},
  {Mizuno}, {Hardee}, {Niemiec}, {G{\'o}mez}, {Pe'er}, {Du{\c{t}}an}, {Meli},
  {Sol}, {Pohl}, \& {Hartmann}}]{Nishikawa_16}
{Nishikawa}, K.~I., {Frederiksen}, J.~T., {Nordlund}, {\r{A}}., {et~al.} 2016,
  \apj, 820, 94, \dodoi{10.3847/0004-637X/820/2/94}

\bibitem[{Osmanov {et~al.}(2008)Osmanov, Mignone, Massaglia, Bodo, \&
  Ferrari}]{Osmanov_08}
Osmanov, Z., Mignone, A., Massaglia, S., Bodo, G., \& Ferrari, A. 2008,
  Astronomy \& Astrophysics, 490, 493

\bibitem[{Pausch {et~al.}(2017)Pausch, Bussmann, Huebl, Schramm, Steiniger,
  Widera, \& Debus}]{Pausch_17}
Pausch, R., Bussmann, M., Huebl, A., {et~al.} 2017, Phys. Rev. E, 96, 013316,
  \dodoi{10.1103/PhysRevE.96.013316}

\bibitem[{{Petropoulou} \& {Sironi}(2018)}]{petropoulou_18}
{Petropoulou}, M., \& {Sironi}, L. 2018, \mnras, 481, 5687,
  \dodoi{10.1093/mnras/sty2702}

\bibitem[{Prajapati \& Chhajlani(2010)}]{Prajapati_10}
Prajapati, R.~P., \& Chhajlani, R.~K. 2010, Physics of Plasmas, 17, 112108,
  \dodoi{10.1063/1.3512936}

\bibitem[{{Rieger}(2019)}]{rieger_19}
{Rieger}, F.~M. 2019, Galaxies, 7, 78, \dodoi{10.3390/galaxies7030078}

\bibitem[{{Ripperda} {et~al.}(2020){Ripperda}, {Bacchini}, \&
  {Philippov}}]{ripperda_20}
{Ripperda}, B., {Bacchini}, F., \& {Philippov}, A.~A. 2020, \apj, 900, 100,
  \dodoi{10.3847/1538-4357/ababab}

\bibitem[{Ryu {et~al.}(2000)Ryu, Jones, \& Frank}]{Ryu_00}
Ryu, D., Jones, T.~W., \& Frank, A. 2000, The Astrophysical Journal, 545, 475,
  \dodoi{10.1086/317789}

\bibitem[{{Sharma} \& {Chhajlani}(1998)}]{Sharma_98}
{Sharma}, P.~K., \& {Chhajlani}, R.~K. 1998, Physics of Plasmas, 5, 625,
  \dodoi{10.1063/1.872780}

\bibitem[{{Sikora} {et~al.}(2016){Sikora}, {Rutkowski}, \&
  {Begelman}}]{sikora_16b}
{Sikora}, M., {Rutkowski}, M., \& {Begelman}, M.~C. 2016, \mnras, 457, 1352,
  \dodoi{10.1093/mnras/stw107}

\bibitem[{{Sironi} \& {Beloborodov}(2020)}]{sironi_belo_20}
{Sironi}, L., \& {Beloborodov}, A.~M. 2020, \apj, 899, 52,
  \dodoi{10.3847/1538-4357/aba622}

\bibitem[{{Sironi} {et~al.}(2016){Sironi}, {Giannios}, \&
  {Petropoulou}}]{sironi_16}
{Sironi}, L., {Giannios}, D., \& {Petropoulou}, M. 2016, \mnras, 462, 48,
  \dodoi{10.1093/mnras/stw1620}

\bibitem[{{Sironi} \& {Spitkovsky}(2014)}]{ss_14}
{Sironi}, L., \& {Spitkovsky}, A. 2014, \apjl, 783, L21,
  \dodoi{10.1088/2041-8205/783/1/L21}

\bibitem[{{Sobacchi} \& {Lyubarsky}(2018)}]{Sobacchi_18}
{Sobacchi}, E., \& {Lyubarsky}, Y.~E. 2018, \mnras, 473, 2813,
  \dodoi{10.1093/mnras/stx2592}

\bibitem[{{Spitkovsky}(2005)}]{spitkovsky_05}
{Spitkovsky}, A. 2005, in AIP Conf. Ser., Vol. 801, Astrophysical Sources of
  High Energy Particles and Radiation, ed. {T.~Bulik, B.~Rudak, \&
  G.~Madejski}, 345, \dodoi{10.1063/1.2141897}

\bibitem[{{Tolman} {et~al.}(2018){Tolman}, {Loureiro}, \&
  {Uzdensky}}]{Tolman_18}
{Tolman}, E.~A., {Loureiro}, N.~F., \& {Uzdensky}, D.~A. 2018, Journal of
  Plasma Physics, 84, 905840115, \dodoi{10.1017/S002237781800017X}

\bibitem[{{Uzdensky} {et~al.}(2010){Uzdensky}, {Loureiro}, \&
  {Schekochihin}}]{uzdensky_10}
{Uzdensky}, D.~A., {Loureiro}, N.~F., \& {Schekochihin}, A.~A. 2010, Physical
  Review Letters, 105, 235002, \dodoi{10.1103/PhysRevLett.105.235002}

\bibitem[{{Walker} {et~al.}(2018){Walker}, {Hardee}, {Davies}, {Ly}, \&
  {Junor}}]{walker_18}
{Walker}, R.~C., {Hardee}, P.~E., {Davies}, F.~B., {Ly}, C., \& {Junor}, W.
  2018, \apj, 855, 128, \dodoi{10.3847/1538-4357/aaafcc}

\bibitem[{{Werner} \& {Uzdensky}(2017)}]{werner_17}
{Werner}, G.~R., \& {Uzdensky}, D.~A. 2017, \apjl, 843, L27,
  \dodoi{10.3847/2041-8213/aa7892}

\bibitem[{{Werner} {et~al.}(2016){Werner}, {Uzdensky}, {Cerutti}, {Nalewajko},
  \& {Begelman}}]{werner_16}
{Werner}, G.~R., {Uzdensky}, D.~A., {Cerutti}, B., {Nalewajko}, K., \&
  {Begelman}, M.~C. 2016, \apjl, 816, L8, \dodoi{10.3847/2041-8205/816/1/L8}

\bibitem[{Zhang {et~al.}(2009)Zhang, MacFadyen, \& Wang}]{Zhang_09}
Zhang, W., MacFadyen, A., \& Wang, P. 2009, The Astrophysical Journal, 692,
  L40, \dodoi{10.1088/0004-637x/692/1/l40}

\end{thebibliography}
\bibliographystyle{aasjournal}



\end{document}